\documentclass[reprint,aps,prl,superscriptaddress,
amsmath,amssymb]{revtex4-1}
\usepackage{mathtools}
\usepackage{graphicx}
\usepackage{booktabs,multirow}
\usepackage{braket}

\usepackage[
  bookmarks=false,
  colorlinks,
  linkcolor=blue,
  urlcolor=blue,
  citecolor=blue,
  plainpages=false,
  pdfpagelabels,
  final,
  breaklinks=true
]{hyperref}

\begin{document}
\newcommand{\pd}[2]{\frac{\partial #1}{\partial #2}} 
\newcommand{\td}[2]{\frac{d #1}{d #2}} 

\newcommand{\bs}{\boldsymbol}
\newcommand{\bt}{\textbf}
\newcommand{\sech}{\text{sech}}
\newcommand{\erfc}{\text{erfc}}
\newcommand{\bse}{\begin{subequations}}
\newcommand{\ese}{\end{subequations}}
\newcommand{\im}{\text{i}}
\newcommand{\ud}[0]{\mathrm{d}}
\newcommand{\norm}[1]{\left\lVert#1\right\rVert}

\allowdisplaybreaks

\title{Generalized Gaussian beams in terms of Jones vectors}

\author{R. Guti\'{e}rrez-Cuevas}
\email{rgutier2@ur.rochester.edu}
\affiliation{The Institute of Optics and the Center for Coherence and Quantum Optics, University of Rochester, Rochester, NY 14627, USA}
\author{M. R. Dennis}
\affiliation{School of Physics and Astronomy, University of Birmingham, Birmingham B15 2TT, UK}
\author{M. A. Alonso}
\email{miguel.alonso@fresnel.fr}
\affiliation{The Institute of Optics and the Center for Coherence and Quantum Optics, University of Rochester, Rochester, NY 14627, USA}
\affiliation{Aix Marseille Univ, CNRS, Centrale Marseille, Institut Fresnel, UMR 7249, 13397 Marseille Cedex 20, France}

\date{\today}

\begin{abstract}
Based on the operator formalism that arises from the underlying SU(2) group structure, a formula is derived that provides a description of the generalized Hermite-Laguerre Gauss modes in terms of a Jones vector, traditionally used to describe polarization. This identity highlights the relation between these generalized Gaussian beams, the elliptical ray families, and the Majorana constellations used to represent structured-Gaussian beams. Moreover, it provides a computational advantage over the standard formula in terms of Wigner $d$ functions.
\end{abstract}


\maketitle

Hermite-Gauss (HG) and Laguerre-Gauss (LG) modes are among the best-known solutions for free propagating paraxial fields 
\cite{siegman1986lasers}. They correspond to separable Gaussian solutions in Cartesian and polar coordinates, respectively, and constitute complete bases. 
These beams arise naturally as eigenmodes for optical cavities and gradient-index waveguides 
\cite{siegman1986lasers}, and  have been used extensively to study particle trapping, paraxial propagation, and data transmission 
\cite{andrews2008structured,yao2011orbital,rubinsztein-dunlop2016roadmap}. Particularly, LG modes have been a workhorse for investigations of fields possessing orbital angular momentum and their applications 
\cite{yao2011orbital,andrews2012angular}.

HG and LG beams are said to be self-similar because as they propagate their intensity profile is preserved up to a scaling factor. Moreover, any superposition of HG and LG modes that accumulate the same Gouy phase upon propagation produces also a self-similar beam 
\cite{alonso2017ray,gutierrez-cuevas2019modal}. The amount of Gouy phase gained by HG and LG modes depends linearly on their total order $N$ which is determined from the usual Cartesian indices of HG modes or the radial and azimuthal indices of LG modes. Therefore, the subset of LG (or HG) modes with the same total order $N$ can be used as an orthogonal basis for expressing more general self-similar structured-Gaussian (SG) beams of order $N$
\cite{alonso2017ray,gutierrez-cuevas2019modal}. 

\begin{figure}
\centering
\includegraphics[width=.99\linewidth]{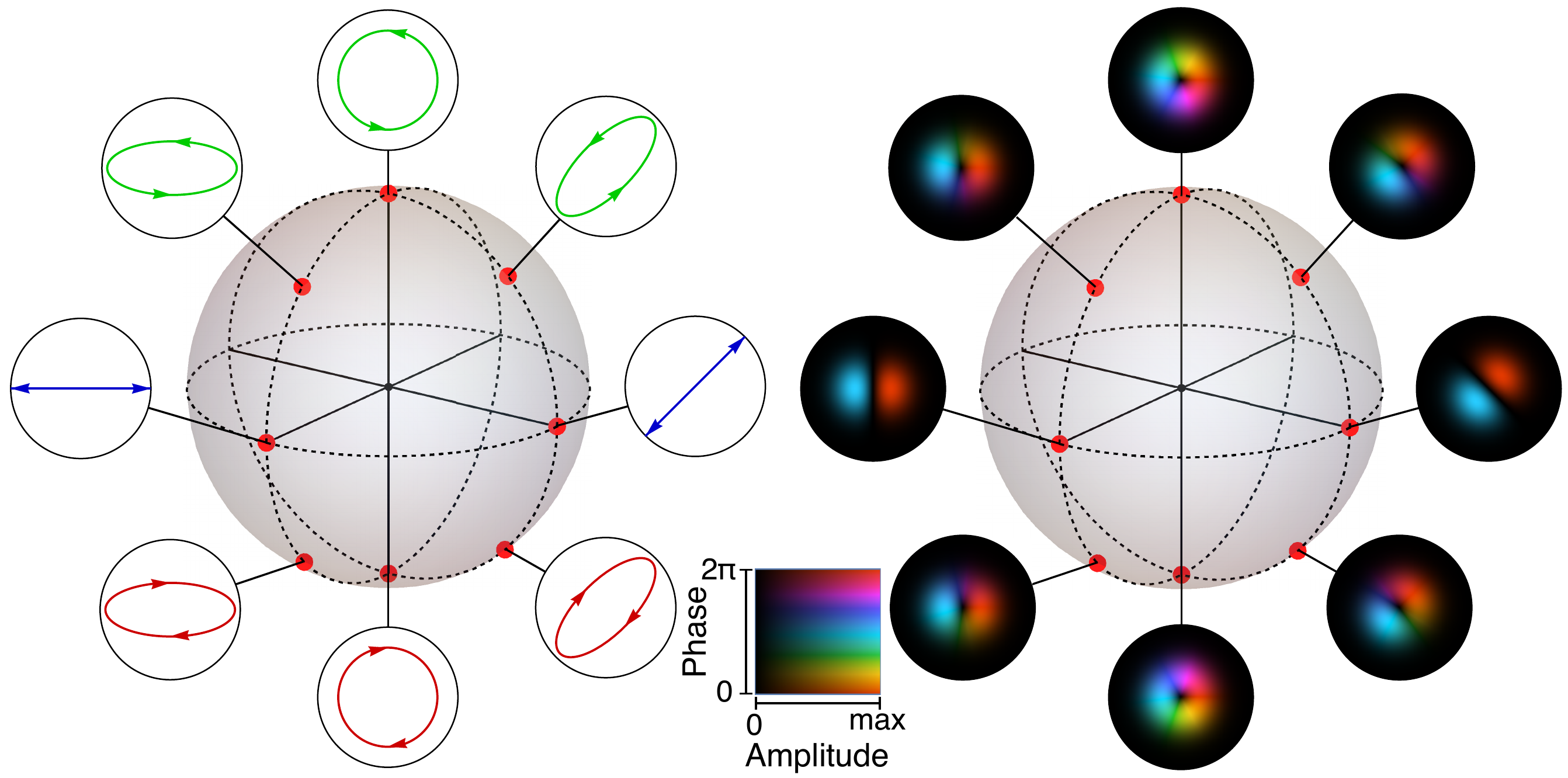}
\caption{\label{fig:PS} (Left) Poincar\'e sphere representing the polarization state of a paraxial field. (Right) Modal Poincar\'e sphere for the lower order ($N=1$) HLG modes.}
\end{figure}

One particularly interesting subfamily of SG beams is that of the generalized Hermite-Laguerre-Gauss (HLG) modes 
\cite{abramochkin2004generalized,deng2008hermitelaguerregaussian,
deng2008elegant} which interpolate between HG and LG modes. 
HLG modes can be obtained experimentally from HG or LG modes via astigmatic transformations implemented with cylindrical lenses   
\cite{abramochkin2004generalized,alieva2007orthonormal}. These modes can be represented as points on the surface of a modal Poincar\'e sphere (MPS), indicating a mathematical analogy between the modal structure of these scalar beams and paraxial polarization  \cite{enk1993geometric,padgett1999poincare,calvo2005wigner,habraken2010universal}, as shown in Fig.~\ref{fig:PS}. 

The aim  of this letter is to further strengthen 
the mathematical analogy between HLG modes and paraxial polarization by showing that the former is also intrinsically linked to a complex 2D ``Jones'' vector. This endeavor is warranted by the fact that it was through the mathematical analogy between the polarization Jones vectors and the first-order SG modes that the Poincar\'e sphere (PS) was first adapted to represent the modal structure of paraxial beams 
\cite{enk1993geometric,padgett1999poincare}. Several PS representations for higher-order beams have been proposed  
\cite{calvo2005wigner,habraken2010universal,
milione2011higher,dennis2017swings,
alonso2017ray,gutierrez-cuevas2019modal} that highlight the mathematical similarity between modal structure and polarization. However, this connection is not reflected by the standard formula of  HLG modes  in terms of Wigner $d$ functions [see Eq.~(\ref{eq:gginlg})]. 
We show in what follows, based on the ray and operator formalisms, that a more intuitive and compact expression exists for HLG scalar beams which clearly highlights the analogy with polarization through the use of Jones vectors. 

The first connection between HLG modes and polarization was made in 
\cite{enk1993geometric} within the context of the Pancharatnam-Berry phase  arising from a cyclic transformation  
\cite{pancharatnam1956generalized,berry1984quantal,malhotra2018measuring}. Yet, it took some years for the geometric PS construction, where the presence of a geometric phase becomes evident,  to be adapted for the description of the first-order ($N=1$) HLG modes 
\cite{padgett1999poincare}. Figure \ref{fig:PS} shows the PS for polarization as determined by the Jones vector,
\begin{align}
\label{eq:jones}
\bt v(\theta, \phi) = \cos \left( \frac{\theta}{2} \right) e^{-\im \frac{ \phi}{2}} \bs \epsilon_+ + \sin \left( \frac{\theta}{2} \right)e^{\im \frac{ \phi}{2}} \bs \epsilon_-,
\end{align}
where, in the context of polarization,  $\bs \epsilon_\pm =(\bt x\pm \im \bt y)/2^{1/2}$ represent circularly polarized light.  Throughout this work,  $0 \leq \theta \leq \pi$ and $0 \leq \phi \leq 2\pi$ denote the polar and azimuthal angles, respectively, for all the PS treated. For shorthand, the unit vector $\bt u =(u_1,u_2,u_3)= (\cos \phi \sin \theta, \sin \phi \sin \theta, \cos \theta)$ is used to denote points on the surface of a sphere. 
Note that the normalized 2D complex Jones vector $\bt v$ and the real 3D unit vector $\bt u$ both encode points on the surface of the MPS through their dependence on the angles $\theta$ and $\phi$. 
The first-order HLG modes can be described by a similar expression,
\begin{multline} 
\label{eq:gg1}
\text{GG}_{1,1}(\bt u ;\bt r) = \cos \left( \frac{\theta}{2} \right) e^{-\im \frac{ \phi}{2}} \text{LG}_{1,1}(\bt r)\\ + \sin \left( \frac{\theta}{2} \right)e^{\im \frac{ \phi}{2}} \text{LG}_{1,-1}(\bt r),
\end{multline}
where the LG beams play the role of circular polarization thus leading to the modal PS (MPS) shown in Fig.~\ref{fig:PS}. (The index notation for LG beams is explained below.) In these two cases any point on the PS can be expressed as a simple linear combination of the two states represented by the poles. The particular choice of (polarization and modal) states used in Eqs.~(\ref{eq:jones}) and (\ref{eq:gg1}) is, to some extent, arbitrary. Any pair of antipodal points could have been used in their stead since it is their orthogonality that is key and not the states themselves. As we will see this observation remains valid for higher-order modes and permeates into our main result [see Eq.~(\ref{eq:uHLG})]. For this reason, when the unit vector $\bt u $ corresponds to the Jones vector $\bt v$, then $-\bt u $ corresponds to  $\bar{\bt v}(\theta, \phi) = \bt v (\pi- \theta, - \phi)$.

%

\begin{figure}
\centering
\includegraphics[width=.99\linewidth]{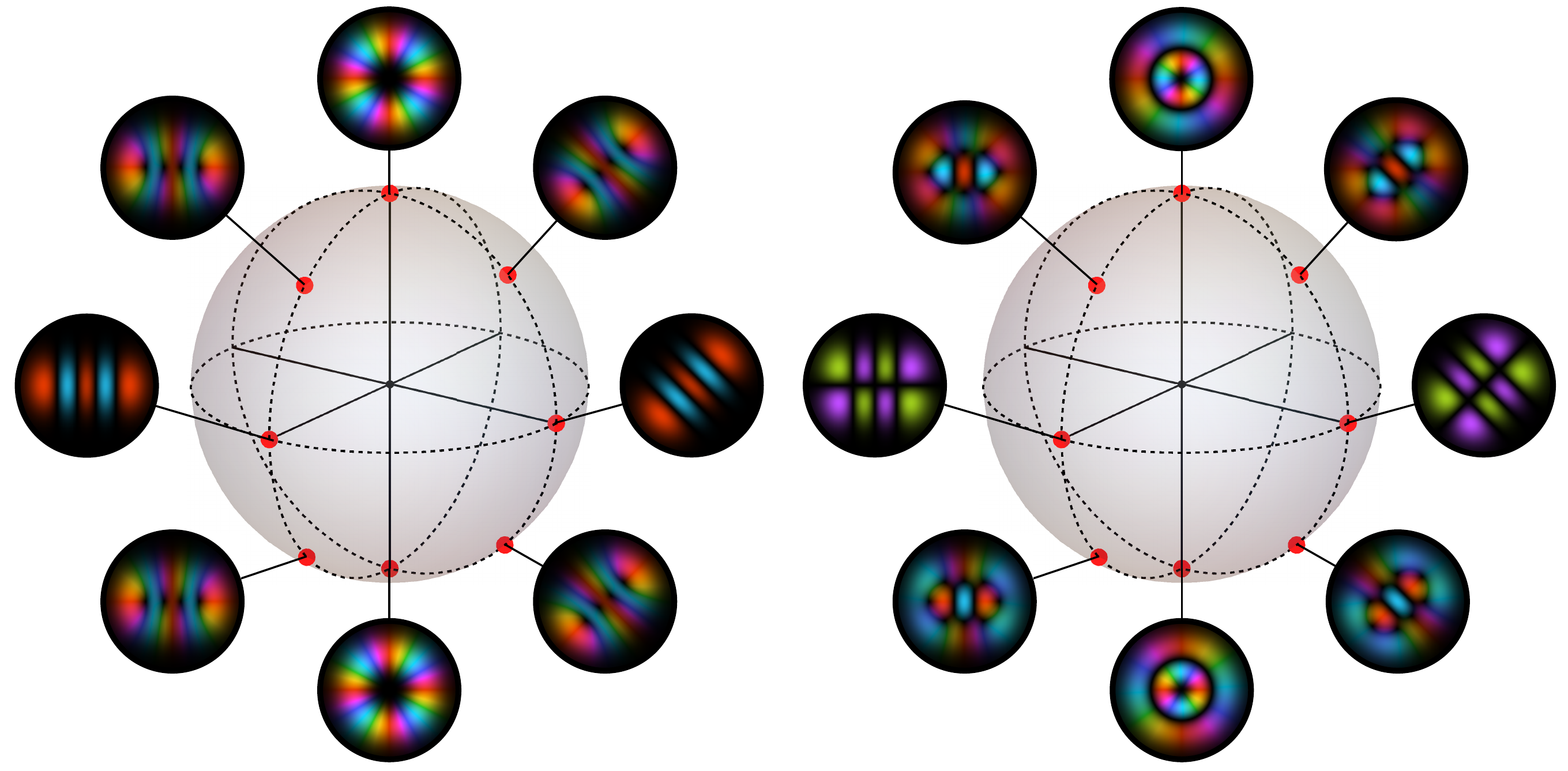}
\caption{\label{fig:MPS4} Modal Poincar\'e sphere for (left) $N=4$, $\ell =4$  and  (right) $\ell =2$. Also shown are the intensity distributions with the phase coded in hue for several HLG beams with their corresponding modal spot.}
\end{figure}

The geometric  representation provided by the PS can be extended to higher-orders modes 
\cite{calvo2005wigner,habraken2010universal}, where the two poles correspond to LG modes of opposite vorticity, the equator corresponds to rotated HG modes and all other points correspond to HLG modes connected via astigmatic transformations. 
This construction is supported by the formula giving the HLG modes in terms of LG modes 
\cite{danakas1992analogies,allen1992orbital,kimel1993relations,
dennis2017swings,gutierrez-cuevas2019modal},
 \begin{align}
\label{eq:gginlg}
\text{GG}_{N,\ell}(\bt u ;\bt r)= \sum_{\frac{\ell'}{2}=-\frac{N}{2}}^{N/2}  e^{-\frac{\im}{2}\ell' \phi}d_{\frac{\ell'}{2},\frac{\ell}{2}}^\frac{N}{2}(\theta) \text{LG}_{N,\ell'}(\bt r),
\end{align}
where  $d_{m',m}^j(\theta)$ is the  Wigner $d$ function and the transverse field profile for the LG mode at its waist plane is given by 
\begin{multline}
\label{eq:csLG}
\text{LG}_{N,\ell}(\bt r)= \frac{\im^{|\ell|-N}}{w}\sqrt{\frac{2^{|\ell|+1}\left(\frac{N-|\ell|}{2}\right)!}{\pi\left(\frac{N+|\ell|}{2}\right)!}}e^{-\frac{r^2}{w^2}}
  \\ \times
 \left( \frac{r}{w}\right)^{|\ell|} e^{\im \ell \varphi} L_{\frac{N-|\ell|}{2}}^{|\ell|}\left( \frac{2r^2}{w^2} \right).
\end{multline}
An extra phase factor was introduced in the definition of the LG beams to guarantee that they satisfy the Condon-Shortley convention 
\cite{sakurai2010modern,dennis2017swings,dennis2019gaussian,gutierrez-cuevas2019modal}. 
Note that here we label these modes by the total order $N$ and the azimuthal index $\ell$, with  $\ell$ ranging from $-N$ to $N$ in steps of two. These  are related to the usual radial $p$ and azimuthal $\ell$ indices used to denote LG beams  via $N=2p +|\ell|$. They can also be related to the Cartesian indices $m$ and $n$ along $x$ and $y$, respectively, used for HG beams through $N=m+n$ and $\ell=m-n$.

That is, for given $N$, $\ell$, HLG modes are still uniquely determined by the coordinates $\theta$ and $\phi$ over the MPS. 
In particular, HG modes are given by
\begin{align}
\text{HG}_{N,\ell}(\bt r)=\sum_{\frac{\ell'}{2}=-\frac{N}{2}}^{N/2} d_{\frac{\ell'}{2},\frac{\ell}{2}}^\frac{N}{2}(\pi/2) \text{LG}_{N,\ell'}(\bt r),
\end{align} 
which include Condon-Shortley phases.
Thus, HLG modes interpolate between HG and LG modes with the same $N$ and $\ell$ through astigmatic transformations \cite{allen1992orbital,alieva2007orthonormal}, and
a different MPS is needed for each pair of indices $N$ and $\ell$, as shown in Fig.~\ref{fig:MPS4} for $N=4$ with $\ell=4$ and $\ell=2$. The point ${\bf u}$ that represents a mode is henceforth referred to as the modal spot (MS).
This geometric representation, however, is limited to the description of HLG modes. Note also that for $N>1$ the HLG modes do not correspond to a simple linear combination of the two modes represented by the poles, but to a linear combination of all HG (or LG) modes with the same total order $N$.

\begin{figure}
\centering
\includegraphics[width=.99\linewidth]{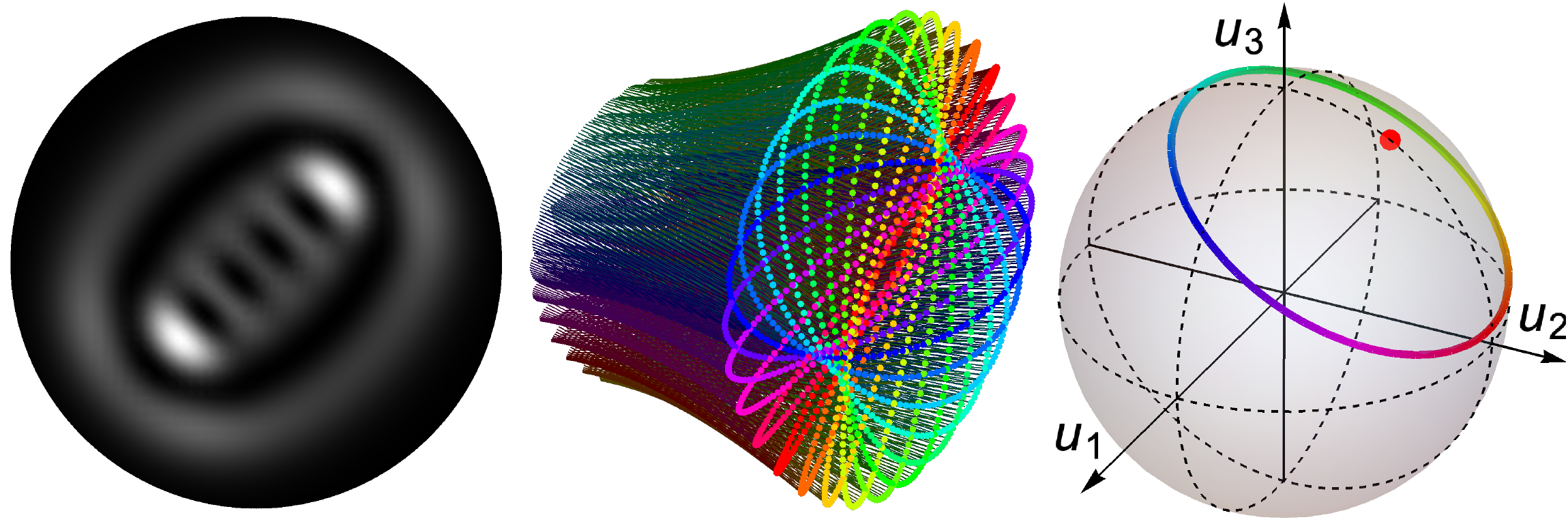}
\caption{\label{fig:wNr} (From left to right) Intensity distribution,  elliptic-ray families, and Poincar\'e path (PP) for the HLG beam of order $N=6$ and $\ell =4$ with $\theta=\pi/6$ and $\phi =\pi/2$. The color of the ellipses of rays corresponds to that of the PP. }
\end{figure}

Another analogy with polarization arises in the context of semiclassical estimates for SG beams  
\cite{dennis2017swings,alonso2017ray,dennis2019gaussian}. Because they are self-similar, the two-parameter family of rays used to describe SG beams is given by an ensemble of ruled hyperboloids whose cross section at any plane of constant $z$ is an ellipse. At the waist plane, it turns out that a convenient parametrization of the transverse position $\bt Q =(Q_x,Q_y)$ and direction $\bt P =(P_x,P_y)$ of the rays within each hyperboloid is through a Jones vector via
\begin{align} \label{QandP}
\bt q + \im \bt p = \sqrt{N+1} \bt v (\theta,\phi )e^{-\im \tau},
\end{align}
where $\bt Q= w \bt q$ and $\bt P =2 \bt p / k w$.
As $\tau$ varies, both $\bt Q$ and $\bt P$ trace an ellipse which, in analogy with polarization, can be represented by a point on a ray PS (RPS) 
\cite{alonso2017ray,dennis2019gaussian,malhotra2018measuring}. The ensemble of elliptic ray families (i.~e.~ruled hyperboloids) describing a SG beam corresponds to a Poincar\'e path (PP)
on the RPS. For HLG beams the PP corresponds to a circle centered at the MS and whose radius depends on $\ell$ (with higher $|\ell|$ corresponding to smaller radii). Figure \ref{fig:wNr} shows the PP, the rays and the corresponding intensity profile at the waist for a HLG mode. This construction enables the representation of more general SG beams. Moreover, it allows SG beams with the same total $N$ to coexist in the same PS, thus solving the limitations of the MPS.

Dressing the rays with appropriate Gaussian contributions leads to a semiclassical estimate that is exact for HLG modes 
\cite{alonso2017ray,dennis2019gaussian}. After integrating in $\tau$, this estimate takes the form of a continuous superposition of extremal ($\ell=N$) HLG beams which can be identified as the coherent states in the reduced space of SG beams with total order $N$ 
\cite{gutierrez-cuevas2019modal}. In \cite{alonso2017ray} these were expressed as a 
complex-valued HG beam of the form 
\begin{align}
\label{eq:uGG}
 \text{GG}_{N,N}(\bt v ; \bt r) =& (\bt v \cdot \bt v) ^{N/2} U_N(\bt v ; \bt r)U_0( \bar{\bt v}; \bt r),
\end{align}
where 
\begin{align}
\label{eq:un}
U_j(\bt v; \bt r)
=& \sqrt{\frac{1}{w \sqrt{\pi} \; 2^{j-\frac 1 2} j!}}   e^{-\frac{r^2}{2 w^2}}H_j\left( \frac{\sqrt 2 \; \bt v \cdot \bt r}{w\sqrt{\bt v \cdot \bt v}} \right),
\end{align}
Notice that we used the identity $\bt v \cdot \bt v = \sin \theta$.
The expression in Eq.~(\ref{eq:uGG}) was later formally proven to be the extremal ($\ell=N$) HLG beam 
\cite{dennis2019gaussian}.
This complex-valued HG expression was independently derived in 
\cite{kotlyar2015vortex} and referred to as a vortex HG beam, but the connection to the Jones vector was not made. 
The expression given by Eqs.~(\ref{eq:uGG}) and (\ref{eq:un}) is much simpler since it is composed of a single term, whereas the standard one given in Eq.~(\ref{eq:gginlg}) has $N+1$ terms. More importantly, Eq.~(\ref{eq:uGG}) provides a connection to the ray description through the Jones vector which is made explicit in the arguments of the HLG by using $\bt v$ to denote the location of the corresponding HLG beam on the surface of the MPS, instead of $\bt u$. 
The goal of this work is to find similar expressions for all other HLG modes.

HG and LG beams can also be described through an operator formalism analogous to that of the two-dimensional isotropic oscillator 
\cite{danakas1992analogies,enk1992eigenfunction,
nienhuis1993paraxial,calvo2005wigner,
habraken2010universal,dennis2017swings,dennis2019gaussian,gutierrez-cuevas2019modal}. It was actually for this system that the coherent states in Eqs.~(\ref{eq:uGG}) and (\ref{eq:un}) were first derived \cite{pollett1995elliptic}. Using Schwinger's oscillator model  \cite{sakurai2010modern}, HG and LG modes can be described as eigenfunctions of the operators \cite{nienhuis1993paraxial,dennis2017swings,gutierrez-cuevas2019modal}
\bse
\begin{align}
\widehat{T}_1 =&\frac{1}{2w^2} ( x^2- y^2) -\frac{w^2}{8}\left( \pd{^2}{x^2} - \pd{^2}{y^2} \right), \\
\widehat{T}_2 =&\frac{1}{w^2} x y -\frac{w^2}{4}  \pd{^2}{x\partial y} , \\
\widehat{T}_3 =&-\frac{\im}{2} \left( x  \pd{}{y} -  y  \pd{}{x} \right) ,
\end{align}
\ese
which satisfy the commutation relation for quantum angular momentum,
\begin{align}
[\widehat{T}^2,\widehat T_j]=0, \qquad [\widehat T_i,\widehat T_j]= \im \sum_k \epsilon_{ijk} T_k,
\end{align}
with $ \epsilon_{ijk}$ being the Levi-Civita tensor. 

The standard convention is to  take  $\widehat{T}_3$ and its eigenfunctions (the LG modes) as a reference, but this choice is arbitrary. The most general situation is to consider rotated versions of the operators $\widehat{T}_j$: 
\bse
\begin{align}
\widehat{T}_1 (\bt u) =& \cos \theta \cos \phi \widehat{T}_1 + \cos \theta \sin \phi \widehat{T}_2 - \sin \theta \widehat{T}_3, \\
\widehat{T}_2 (\bt u) =& - \sin \phi \widehat{T}_1 + \cos  \phi \widehat{T}_2 , \\
\widehat{T}_3(\bt u) =&\sin \theta \cos \phi \widehat{T}_1 + \sin \theta \sin \phi \widehat{T}_2 + \cos \theta \widehat{T}_3,
\end{align}
\ese
These operators satisfy the same commutation relation as their unrotated versions. 
Moreover, the eigenvalue equation 
\begin{align}
\label{eq:eigGG}
\widehat{T}_3(\bt u) \text{GG}_{N,\ell}(\bt u ; \bt r)  =\frac{ \ell}{2} \; \text{GG}_{N,\ell}(\bt u ; \bt r)  
\end{align}
is satisfied  \cite{dennis2017swings}. If the HLG beam satisfying Eq.~(\ref{eq:eigGG}) is taken as a reference then  annihilation and creation operators can be defined as
\begin{align}
\widehat{T}_\pm (\bt u) = &\widehat{T}_1 (\bt u) \pm \im \widehat{T}_2 (\bt u),
\end{align}
which satisfy
\begin{multline}
\widehat{T}_\pm (\bt u) \text{GG}_{N,\ell}(\bt u ; \bt r) \\
=\frac{1}{2}\sqrt{(N\mp \ell)(N\pm \ell+2)}\; \text{GG}_{N,\ell\pm 2}(\bt u ; \bt r) ,
\end{multline}
where the Condon-Shortley convention was assumed.

Starting from the extremal HLG mode, $\text{GG}_{N,N}$, one can then find all other modes by successive  application of the appropriate annihilation operator. Particularly, using the expressions given in Eqs.~(\ref{eq:uGG}) and (\ref{eq:un}), and several mathematical identities involving the Hermite polynomials, it can be shown that 
\begin{multline}
\widehat{T}_- (\bt u) U_m \left( \bt v ; \bt r\right) 
U_n \left(\bar{\bt v} ; \bt r \right)\\
=  -\im \sqrt{m(n+1)}U_{m-1} \left( \bt v; \bt r\right) 
U_{n+1} \left( \bar{\bt v} ; \bt r \right)  \\
 - \cot \theta \sqrt{m(m-1)} U_{m-2} \left( \bt v; \bt r\right) 
U_{n} \left(\bar{\bt v} ; \bt r \right) ,
\end{multline}
from which it follows that
\begin{align}
\label{eq:uHLG}
\text{GG}_{N,\ell}(\bt v ; \bt r) 
=& \sum_{j=0}^{\frac{N-\ell}{2}} (- \im )^{\frac{N-\ell}{2}+j} \sqrt{ {{\frac{N+\ell}{2}}\choose{j}} {{\frac{N-\ell}{2}}\choose{j}}  } \cos^j \theta \nonumber \\
&\times  \sin^{\frac{N}{2}-j} \theta  U_{\frac{N+\ell}{2}-j} \left( \bt v ; \bt r\right) 
U_{\frac{N-\ell}{2}-j} \left( \bar{\bt v} ; \bt r \right) .
\end{align}
This equation is the main result of this work.  It is worth pointing out that Eq.~(\ref{eq:uHLG}) can be anticipated from the work in  \cite{wuensche2001hermite} which relates two-dimensional extensions of Hermite and Laguerre polynomials.
The first thing to notice is that this expression has only $(N-\ell)/2+1$ terms, which is considerably fewer than the $N+1$ terms used in the standard expression in  Eq.~(\ref{eq:gginlg}). The maximum number of terms required in Eq.~(\ref{eq:uHLG}) is $\lfloor N/2 + 1 \rfloor$, since for $\ell<0$ the same expression  can be evaluated at the antipodal point, $-\bt u$. Thus, this formula provides a computational advantage. More importantly,
the reason for this compactness is that the expression is written in terms of the Jones vectors of the modal spot and its antipode. 
Furthermore, 
it can be shown through the use of recurrence relations for Hermite polynomials that the function $U_j$ in Eq.~(\ref{eq:un}) satisfies 
\begin{align}
\label{eq:recU}
\sqrt{j+1} U_{j+1}(\bt v; \bt r)= \frac{2 \bt v \cdot \bt r}{\sqrt{\sin \theta}}U_{j}(\bt v ; \bt r)
-\sqrt{j} U_{j-1}(\bt v ; \bt r),
\end{align}
from which the recursion formulas given in 
\cite{abramochkin2004generalized,alieva2005mode,
alieva2007orthonormal} can be rewritten in terms of Jones vectors as
\bse
\begin{align}
2 \bt v \cdot \bt r  \; \text{GG}_{N,\ell}(\bt v ; \bt r)  
=&\sqrt{N+\ell+2} \;\text{GG}_{N+1,\ell+1}(\bt v ; \bt r)  \nonumber\\
&+\sin \theta \sqrt{N+\ell}\; \text{GG}_{N-1,\ell-1}(\bt v ; \bt r) \nonumber \\
&+\cos \theta \sqrt{N-\ell} \;\text{GG}_{N-1,\ell+1}(\bt v ; \bt r), 
\end{align}
\begin{align}
 -\im 2  \bar{\bt v}\cdot \bt r  \; \text{GG}_{N,\ell}(\bt v ; \bt r)  
=&\sqrt{N-\ell+2} \;\text{GG}_{N+1,\ell-1}(\bt v ; \bt r)  \nonumber\\
&+\cos \theta \sqrt{N+\ell}\; \text{GG}_{N-1,\ell-1}(\bt v ; \bt r) \nonumber \\
&-\sin \theta \sqrt{N-\ell}\; \text{GG}_{N-1,\ell+1}(\bt v ; \bt r) .
\end{align}
\ese
It is worth pointing out that the right-hand side of these equations is written purely in terms of $\bt v$.

\begin{figure}
\centering
\includegraphics[width=.99\linewidth]{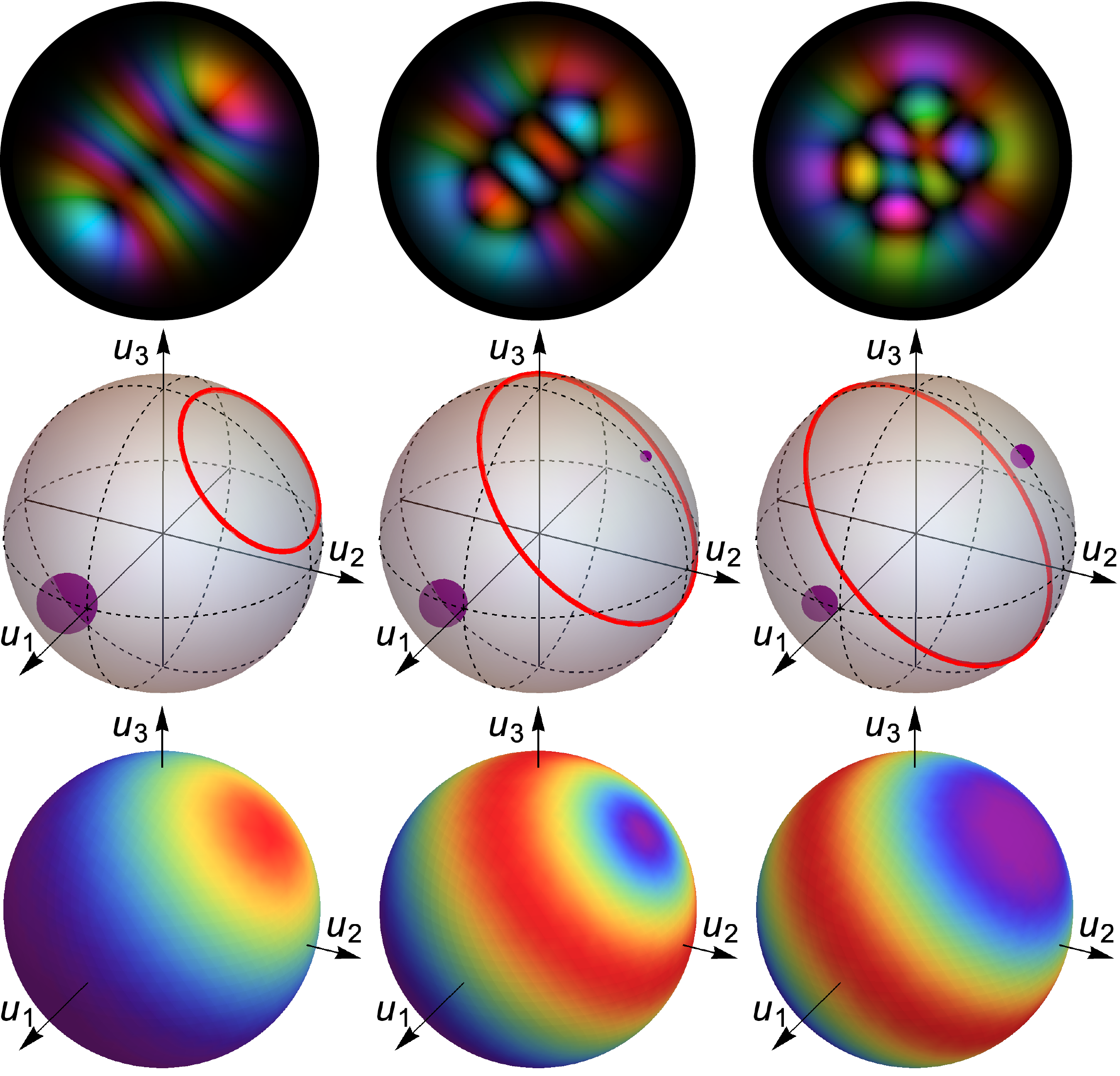}
\caption{\label{fig:hlgMC} (First row) Intensity distribution with the phase coded in hue, and (second row)  Majorana constellation, with the size indicating the number of stars, and the PP and (third row) the $Q$ function for the HLG beams of order $N=5$ along $\theta=\pi/4$ and $\phi =\pi/2$ with (from left to right) $\ell=5,3,1$. }
\end{figure}

Another interesting feature emerges form the expression in Eq.~(\ref{eq:uHLG}): as $\ell$ decreases, the order of the polynomial contributions at the modal spot $\bt u$ decreases while those at the antipodal point increase. This behavior is reminiscent of that of the corresponding Majorana constellations for HLG beams 
\cite{majorana1932atomi,penrose1996shadows,bacry1974orbits,dennis2004canonical,
gutierrez-cuevas2019modal}. As was shown in 
\cite{gutierrez-cuevas2019modal} the proper way to extend the MPS to higher-order modes is through the Majorana representation, originally proposed for spin systems, where a general SG beam is represented by $N$ points (or stars) on the surface of the modal Majorana sphere.
For the particular case of the HLG modes, the Majorana constellation is composed of $(N-\ell)/2$ stars at the modal spot and $(N+\ell)/2$ at the antipodal point (note that there are more stars at the antipodal point than at the modal spot). This is shown in Fig.~\ref{fig:hlgMC} for several HLG modes. With each application of the annihilation operator $\ell$ decreases by two (remember that $\ell$ changes in steps of two) and a star moves from the antipodal point to the modal spot. This behavior is reflected in Eq.~(\ref{eq:uHLG}) where the maximum order of the polynomials exactly corresponds to the number of stars in the constellation. 

The Majorana representation is related to the RPS through the Q (or Husimi) function \cite{alonso2017ray,bengtsson2017geometry,gutierrez-cuevas2019modal}. Each representation uses different features of the Q function to describe a SG beam: the stars correspond to its zeros and the PP corresponds approximately to the ridge outlined by the regions of maximum intensity. This can be seen for the particular case of the HLG beams in Fig.~\ref{fig:hlgMC} where the Q function is shown along with the corresponding MC and PP. Note that, even though both of these representations are significantly more general than the standard MPS, the Majorana representation is the most general since it can be used to describe any SG beam without requiring a well-defined PP (see for example the beams introduced in \cite{gutierrez-cuevas2019modal}).

To summarize, the expression for the HLG beams in Eq.~(\ref{eq:uGG}) clearly highlights the mathematical analogy between these modes and polarization through the use of Jones vectors. 
This form also provides direct connections with more general geometrical representations of SG beams over a Poincar\'e sphere: the ray-based PP and the Majorana constellation, which provide complementary pictures by describing different features of the Q function (its maxima and zeros, respectively). The expression derived for HLG modes is not only more connected to intuition but also more compact, thus providing a computational advantage by reducing the terms to less than half of those used in the standard formula in terms of the Wigner $d$ function. 
The results presented here reveal further structure of the MPS for structured light by finding connections through the underlying SU(2) structure.
\\

\bt{Funding.} National Science Foundation (NSF) (PHY-1507278); the Excellence Initiative of Aix-Marseille University - A*MIDEX, a French ``Investissements d'Avenir'' programme.


%

\end{document}